\documentclass[twocolumn,prb,superscriptaddress,showpacs]{revtex4}
\usepackage{epsfig}
\usepackage{times}
\usepackage{amsmath}

\begin{document}

\title{Pairing states of a polarized Fermi gas trapped in a one-dimensional optical lattice}

\author{A.~E.~Feiguin}
\affiliation{Microsoft Station Q, University of California, Santa Barbara, California 93106}
\author{F. Heidrich-Meisner}
\affiliation{Materials Science and Technology Division, Oak Ridge National Laboratory, Oak Ridge, Tennessee, 37831, and Department of Physics and Astronomy, University of Tennessee, Knoxville, Tennessee 37996, USA}

\date{\today}

\begin{abstract}
We study the properties of a one-dimensional (1D) gas of fermions trapped 
in a lattice by means of the density matrix renormalization group method, focusing on the case of unequal spin populations, and strong attractive interaction.
In the low density regime, 
the system phase-separates into a well defined superconducting core and a fully polarized metallic 
cloud surrounding it. We argue that the superconducting phase corresponds to a 1D
analog of the Fulde-Ferrell-Larkin-Ovchinnikov (FFLO) state, with a quasi-condensate of tightly bound 
bosonic pairs with a finite center-of-mass momentum that scales linearly with the magnetization. 
In the large density limit, the system allows for four phases: in the core, we either find 
a Fock state of localized pairs or a metallic shell with free spin-down fermions moving in a fully filled background of 
spin-up fermions. As the magnetization increases, 
 the Fock state disappears to give room for a metallic phase, with a partially polarized superconducting FFLO shell and a fully polarized metallic cloud surrounding the core.
\end{abstract}

\maketitle

Ultracold atoms in optical lattices can be used to study models of strongly correlated fermions in
clean and controlled experimental conditions. In particular, cold gases provide an optimal playground
to study the crossover between a Bardeen-Cooper-Schrieffer (BCS) superfluid, with extended Cooper pairs,
and a Bose-Einstein condensate (BEC), composed of molecules of tightly bound pairs.\cite{bloch_review}
 As cold atom gases can also be realized in optical lattices, dimensional crossover effects can be accessed.\cite{crossover Stoferle PRL 04}
 In particular, 1D optical lattices can be prepared by strongly ramping up the amplitudes
 of two out of three counterpropagating light waves.\cite{crossover Stoferle PRL 04}
 This, combined with the possibility of tuning the  
 interactions, allows for the realization of the fermionic 1D Hubbard model, with, in experiments, two hyperfine states interacting via an onsite potential.\cite{Moritz PRL 05}
 In the case of a spin imbalanced fermion mixture,  magnetized superconducting states are expected, such as, e.g., a superfluid-normal mixture\cite{Caldas} or the one predicted by
Fulde and Ferrell\cite{FF} and Larkin and Ovchinnikov\cite{LO} (FFLO state) over four decades ago.
The FFLO state\cite{Leo AOP07} is characterized by pairing across a spin-split Fermi surface,
with the resulting Cooper pairs having a finite center-of-mass momentum, proportional to the spin
polarization, and  consequently, an oscillatory phase in the superconducting correlation function.
Its experimental observation has eluded condensed matter physicist until very recently,
 when it was detected in heavy-fermion systems.\cite{Radovan Nature 03}
Its realization in cold atom systems has acquired particular relevance for the field
of high-$T_c$ superconductivity. Moreo and Scalapino \cite{Moreo and Scalapino PRL07} have recently pointed out
 that, by exploiting a particle-hole transformation,\cite{Emery}
its presence in the 2D attractive Hubbard model may imply
the existence of a striped phase in its repulsive counterpart. 

Experiments on 3D traps \cite{ultracold1,ultracold2}
suggest the existence of phase-separated shells, with a superfluid core, and a partially polarized normal cloud, and a future characterization of the superconducting phases 
may establish whether the FFLO phase is present or not.
Theoretical work indicates that the FFLO phase is stable in a narrow window around the (unpolarized) Fermi surface.\cite{Leo AOP07}
However, the instability against an FFLO state may be enhanced in low dimensions, rendering this state more robust.

The uniform 1D polarized Fermi gas has been studied by 
means of bosonization and renormalization group techniques,\cite{Yang PRB01} providing evidence for the  existence 
of an FFLO state. 
The case of fermions confined to a 1D trap has been analyzed using modified versions of the Gaudin-Yang Hamiltonian,\cite{Gaudin 67,Yang 67}
a minimal integrable model of fermions in the continuum interacting via a contact potential. A two-shell structure has been predicted,\cite{Orso 07, Hu PRL07}
with a partially polarized phase of the FFLO type in the center of the trap, and either fully paired or fully polarized wings, 
depending on the strength of the magnetic field or, equivalently, the total magnetization.

In this work we investigate the FFLO state in a Hubbard chain, thus accouting for the optical lattice:
\begin{eqnarray}
H =  & - & t \sum\limits^{L-1}_{i=1, {\sigma}} \left(c^\dagger_{i\sigma} c_{i+1\sigma}+h.c.\right)
+ U \sum\limits^L_{i=1} n_{i\uparrow} n_{i\downarrow}\nonumber\\
& + & V\sum\limits^L_{i=1}(x-L/2)^2n_i,
\label{one}
\end{eqnarray}
where $c^\dagger_{\ell\sigma}$ creates a fermion with spin $\sigma=\uparrow,\downarrow$ at 
site $l$, $n_{\ell\sigma}=c^\dagger_{\ell\sigma}c_{\ell \sigma}$, $n_{\ell}=n_{\ell\uparrow}+n_{\ell\downarrow}$ is the local density, 
$t$ is the hopping parameter, and
$U$ is the onsite interaction energy, which in this work is negative. We define $x=ia$, where $a$ is the lattice spacing, set to unity. 
We add a harmonic confining potential parameterized by a constant $V$.
The  Hubbard model  with $V=0$ has been extensively studied, and its properties are 
well documented in the literature.\cite{Hubbard reviews}
 The low-energy properties of the Hamiltonian~(\ref{one}) with $U<0$ are those of a  Luther-Emery liquid.\cite{Luther-Emery} At small attractive 
interactions, fermions form Cooper-pair-like bound states with  a spin gap, reminiscent of the 
superconducting gap in conventional BSC superconductors. In the case of strong interactions, the 
pairs become tightly bound with their extension of the order of the lattice spacing only, 
effectively behaving as hard-core bosons, while the spin degree of freedom moves to high energies. Superconducting correlations then decay algebraically.
The parabolic trap adds an extra ingredient that leads to the emergence of non-uniform states,  
 and the inclusion of a lattice may produce a richer phase diagram. It has been shown, for instance, 
 that this allows for the possibility of engineering states of hard-core bosons such as pure Fock 
 states that cannot be found in the continuum. \cite{Rigol 04}
At the same time, the physics of the system without a lattice can be recovered in the low density limit.
While recent studies have addressed the {\it unpolarized} situation \cite{unpolarized}, this article's chief case is 
the {\it polarized} one. Note that the corresponding situation with 
repulsive interactions has been studied in, e.g., Ref.~\onlinecite{lee}.

 We use the density matrix renormalization group (DMRG) method \cite{dmrg} to obtain 
 the ground state properties of this model in finite systems, for different numbers of 
 particles $N=N_{\uparrow}+N_{\downarrow}$, and values of the total magnetization 
 $S^z=(N_{\uparrow}-N_{\downarrow})/2$.
 In our calculations we choose $L=80$, $U=-8t$, and $V=0.002t$, unless otherwise stated. 
 In the following we describe two typical situations in the low and large density regimes,
 which illustrate the main features of the problem.

\begin{centering}
\begin{figure}
\epsfig {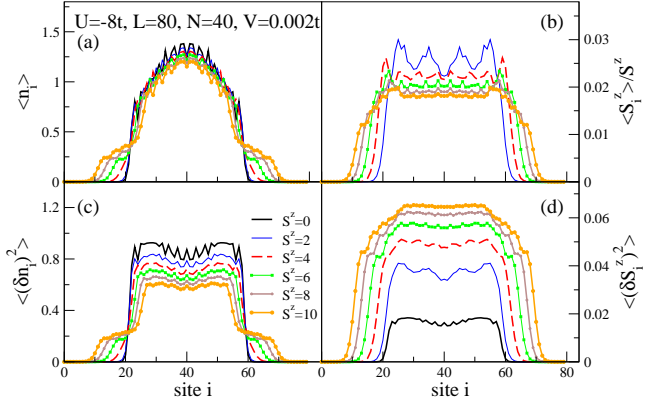}
\caption{
(color online)
Results for the 1D attractive Hubbard model ($U=-8t$, $L=80$), with $N=40$ fermions 
confined by a parabolic potential with  strength $V=0.002t$,  for different magnetizations: 
(a) Density profile, (b) magnetization profile, (c) charge fluctuations, and (d) spin fluctuations. 
}
\label{fig1}
\end{figure}
\end{centering}

We start  by  looking  at the low density case with $N=40$ particles.
In Fig.\ref{fig1} we show the local density $\langle n_i\rangle$ and the 
spin projection $\langle S^z_i\rangle$ along with the fluctuations of both quantities. We define the charge
fluctuations  as $\langle(\delta n_i)^2\rangle=\langle n_i^2\rangle-\langle n_i \rangle^2$, 
and similar for the spin fluctuations $\langle( \delta S^z_i)^2 \rangle$.

The system exhibits a
 nearly half-filled density profile in the center with a sharp edge, 
 and oscillations accompanied by large fluctuations. As  the magnetization grows, 
 the core of the system becomes partially polarized, displaying charge and spin oscillations, 
 surrounded by fully polarized clouds with spin $\uparrow$-fermions only. We can clearly identify two 
 well defined phases: the fully polarized metallic wings, surrounding a polarized state 
 in the core that we wish to characterize in more detail in the following.

\begin{centering}
\begin{figure}
\epsfig {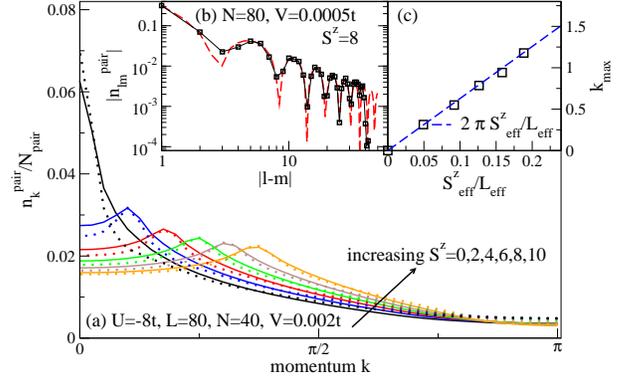}
\caption{
(color online)
(a) Pair momentum distribution function for $N=40$ fermions with $N^{\mathrm{pair}}=\sum_k n_k^{\mathrm{pair}}$. 
The dotted lines show results for the sum in Eq.~(\ref{nkpair}) restricted to the center of the trap.
(b) Spatial decay of pair correlations (squares). Dashed line:
Fit  to $n_{lm}^{\mathrm{pair}} \propto \cos{(k\,|l-m|)}/|l-m|^{\alpha}$ (compare
Ref.~\onlinecite{Yang PRB01}). Data shown for $L=100$, $N=80$, $V=0.0005t$, $S^z=8$,
corresponding to the same effective density $\rho_{\mathrm{eff}}=N\sqrt{V}$.
(c) Momentum $k_{\mathrm{max}}$, at which the distribution shown in (a) is peaked, 
vs. magnetization of the core $S^z_{\mathrm{eff}}/L_{\mathrm{eff}}$ ($L_{\mathrm{eff}}\approx 40$, see text).
}
\label{fig2}
\end{figure}
\end{centering}

 For this purpose, it is illustrating to analyze the natural pair excitations of the system 
 in terms of a bosonic description with creation and annihilation operators 
 $b^\dagger_i=c^\dagger_{i\uparrow}c^\dagger_{i\downarrow}$ and $b_i=c_{i\downarrow}c_{i\uparrow}$. 
  In the low-density limit, indeed, the pairs behave in good approximation like bosons
since $[b^\dagger_i,b_i] = 1-2n_i \approx 1$, 
 but we will use it generically, as it provides an intuitive picture. We further define 
 the pair one-particle density matrix (OPDM) as $\rho_{ij}=\langle b^{\dagger}_i b_j \rangle$. 
 
In Fig.~\ref{fig2}, we show the pair momentum distribution function (MDF) 
\begin{equation}
n_k^{\mathrm{pair}} = (1/L)\sum_{lm} \mbox{exp}[ik(l-m)]\, \rho_{lm}\,. \label{nkpair}
\end{equation} 
 In the unpolarized case, 
we see a sharp peak centered at momentum $k=0$, as observed for quasi-condensates of 
hard-core bosons.\cite{Rigol 04} As the magnetization increases, 
the distribution exhibits two maxima, centered at momenta $\pm k_{\mathrm{max}}$.
Note that by restricting the summation over $l,m$ to the core part of the systems, the
features in $n_k^{\mathrm{pair}}$ become more prominent (dotted lines in Fig.\ref{fig2}).
From Fig.~\ref{fig2}~(c), we see that momentum $k_{\mathrm{max}}$ is proportional to the magnetization in the core $S_{\mathrm{eff}}$, that is obtained by integrating $\langle S^z_i \rangle$ over
 the effective size of the region occupied by the FFLO state, from Fig.\ref{fig1}~(a).
Our results reproduce precisely the behavior expected for the FFLO state, in which the pairs possess a finite center-of-mass 
momentum $k=k_{F\uparrow}-k_{F\downarrow}$ (where $k_{F\sigma}$ is the Fermi vector of the spin-$\sigma$ fermions), 
which is predicted to grow as $k=\pi(n_\uparrow-n_\downarrow)/L=2\pi S^z/L$.\cite{Leo AOP07}
In Fig.\ref{fig2}~(b) we show the spatial decay of the pair correlations, consistent with a power-law decay of the form
$n_{lm}^{\mathrm{pair}} \propto \cos(k_{\mathrm{max}} |l-m|)/|l-m|^{\alpha}$, as predicted in 
Ref.~\onlinecite{Yang PRB01}. 

\begin{centering}
\begin{figure}
\epsfig {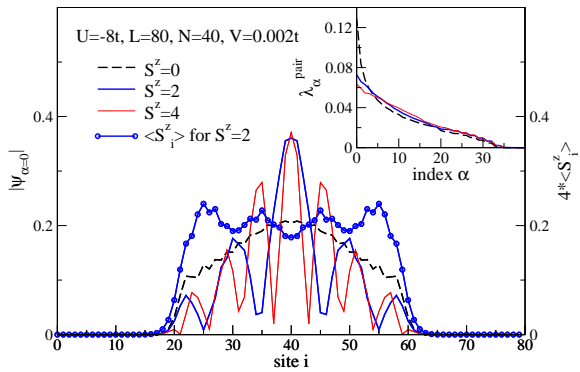}
\caption{
(color online)
Natural orbital $\psi_0$ of the pair OPDM for $N=40$ confined fermions and different magnetizations.
The inset shows the OPDM's eigenvalues $\lambda_\alpha$
}
\label{fig3}
\end{figure}
\end{centering}

The bosonic quasi-condensate can  be studied  by means of Penrose 
and Onsager's description of the superfluid order parameter.\cite{Penrose and Onsager and Leggett}
 The natural orbitals (NO) $\psi_\alpha$ of the system will simply be the single particle 
  eigenstates  -- in the bosonic sense -- of the pair OPDM, and the corresponding 
 eigenvalues $\lambda_\alpha$ represent their occupations. 
 The NO with the largest eigenvalue, $\psi_0$, is the single-particle state in which quasi-condensation 
 takes place.
 The lowest NO and the eigenvalues  of the OPDM are presented in the main panel and the inset of Fig.~\ref{fig3}, respectively. 
 The distribution of eigenvalues qualitatively  resembles the results for a trapped gas of 
 hard-core bosons.\cite{Rigol 04} In the unpolarized case we observe a macroscopic 
 occupation of the lowest eigenstate, corresponding to the order parameter. The effect of increasing the magnetization
  is to decrease the occupation of the bosonic condensate, rendering the  profile of $\lambda_\alpha$ vs $\alpha$ less pronounced. 
 The lowest NO, representing the bosonic order parameter, resembles the density 
 profile in the unpolarized case, see Fig.~\ref{fig1}(a). As the magnetization increases, it develops sharp 
 oscillations, in agreement with the results for the MDF. As observed in Ref.~\onlinecite{Moreo and Scalapino PRL07}, we see that 
 the unpaired fermions accumulate in the nodes of the order parameter, effectively forming magnetic domain walls.
 The single-particle wave function is well confined to the core of the system, 
 an indication of the phase separation between the FFLO state in the center and 
 the fully polarized metallic wings. This observation, i.e., the confinement of the NO to a window 
 of length $L_{\mathrm{eff}}\approx L/2$ further corroborates the use of the effective quantities $L_{\mathrm{eff}}$ and $S^z_{\mathrm{eff}}$
 in the inset of Fig.~\ref{fig2}(b).

\begin{centering}
\begin{figure}
\epsfig {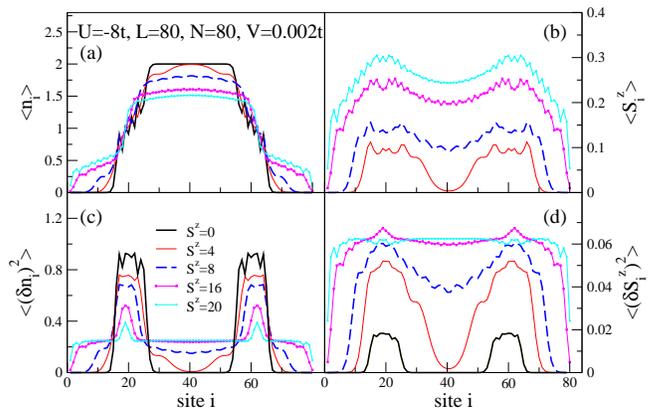}
\caption{
(color online)
Same parameters as in Fig.\ref{fig1}, but $N=80$ fermions.
(a) Density profile, (b) magnetization profile, (c) charge fluctuations, and (d) spin fluctuations.
}
\label{fig4}
\end{figure}
\end{centering}

We now turn our focus to the dense limit, by looking at a system with $N=80$ particles. The results for the local density and magnetization
  are shown in Fig.~\ref{fig4}, along with their fluctuations. 
In the unpolarized case $S^z=0$, a plateau  at density $\langle n_i\rangle=2$ appears in the center, 
surrounded by an extended region with oscillations around $\langle n_i\rangle=1$. The state in the center 
is a pure Fock state of localized pairs, without coherence, and decoupled from the rest of the system. 
As a consequence, the charge and spin fluctuations in the center of the trap are totally suppressed, which are  
signatures of a band insulator with localized bound pairs, and both finite charge and spin gaps. The fluctuations 
increase abruptly at the boundary with the partially-filled, unpolarized region. 
 As the total magnetization increases, pairs are broken and the polarized fermions move toward 
 the edges, effectively suppressing the gaps and the band insulating behavior 
 in the center of the trap. The Fock state survives small values of $S^z$ but disappears rapidly, 
 and the local magnetization becomes almost uniform, roughly equal to the average total magnetization, 
 $\langle S^z_i \rangle \approx S^z/N$. The fluctuations are most prominent in the intermediate region between 
 the wings and the core, and 
after reaching a maximum there, they decrease again to values similar to those observed in the center.

\begin{centering}
\begin{figure}
\epsfig {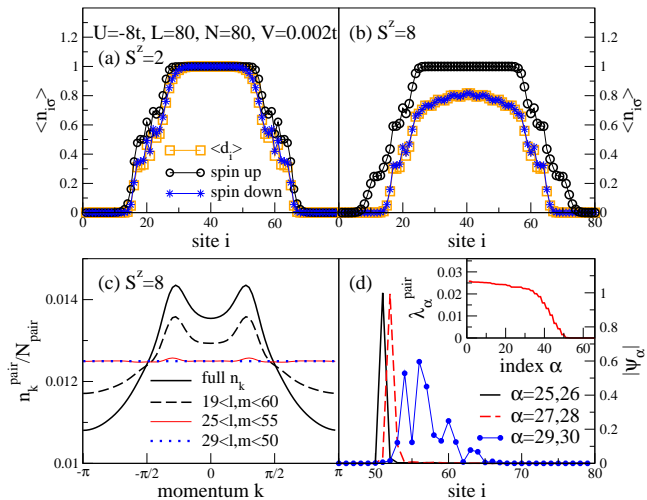}
\caption{
(color online)
(a),(b) Particle density for  $\langle n_{i\sigma}\rangle$, $\sigma=\uparrow,\downarrow$, 
for $N=80$ fermions, and $S^z=2,8$, respectively.  The double occupancy $\langle d_i\rangle=\langle n_{i\uparrow}n_{i\downarrow}\rangle$ 
is also included.
(c) Pair momentum distribution function measured in concentric shells 
of different radii, for $S^z=8$. (d) First three 
NOs of the pair OPDM that are not just localized states.
 The inset shows the eigenvalues of the pair OPDM.
}
\label{fig5}
\end{figure}
\end{centering}

More details are revealed in Figs.~\ref{fig5}(a) and (b), where we plot the densities
 $\langle n_{i\sigma} \rangle$ for $S^z=2$ and $8$, respectively. Obviously, the wings are 
 fully polarized, and the $\uparrow$-spins fill the core of the trap with a plateau of 
 density $\langle n_{i\uparrow} \rangle =1$. The Fock state survives at $S^z=2$ and we find a distribution $\langle n_{i\downarrow} \rangle$
 that resembles the one of trapped spinless fermions.\cite{Rigol 04}
 Fig.~\ref{fig5}(c) shows the MDF calculated in shells centered in the middle of the trap, by restricting the sum 
  in Eq.~(\ref{nkpair}). 
 As we cross the different phases surrounding the core, the distribution evolves from the 
 bulk result with two peaks to a featureless uniform profile describing localized pairs. This shows
 that in the central region, only the $\downarrow$-fermions exhibit any dynamics. 

 These results establish the existence of three distinct phases: 
 first, there is a metallic, fully polarized phase at the wings, 
 behaving like a partially-filled band of spinless fermions  and second, a partially 
 polarized phase in the center that is a metal of freely moving $\downarrow$-fermions in a uniform background of $\uparrow$-fermions, 
 with an effective site energy of $-U$.
 An intermediate, third shell separates the two, with features that resemble those observed 
 in the FFLO state in the low density regime.
 Notice that a fourth phase, the Fock state in the center, 
 survives at small values of the magnetization, before giving way to the metallic core.

In order to shed light on the nature of this intermediate phase, we analyze
the pair OPDM's spectrum and NOs. In Fig.~\ref{fig5}(d), we plot
 the first three NOs that do not correspond to localized bound states. These NOs 
are  precisely situated in the region where the intermediate phase is found. 
We have also calculated the MDF by restricting the summation in Eq.~(\ref{nkpair}) to only this region of the lattice 
(and the reflected sector on the opposite side). This restricted MDF exhibits finite center-of-mass momenta
 that coincide with those observed in Fig.~\ref{fig5}(c) in the full $n_k^{\mathrm{pair}}$, evidencing
  the FFLO phase.   

 Our results in the dilute limit are in general agreement with Refs.~\onlinecite{Orso 07,Hu PRL07}. 
 However, we have not found evidence for a fully paired state in the wings. 
 According to the analysis of Ref.~\onlinecite{Orso 07, Hu PRL07}, this phase should be 
 stable if the magnetic field is smaller than half the binding energy of the bosonic pairs. 
 This should occur below a critical polarization of $P=2S^z/N=0.2$ in the strong coupling limit. 
 In our calculations, for all values of density and magnetization considered (not shown here), 
 we have only seen a fully polarized phase surrounding the core.

In conclusion, we have numerically studied  the attractive Hubbard model 
in a parabolic trapping potential, in different density and magnetization regimes. 
In the unpolarized limit, the system tends to be superconducting, but when the density increases, 
an insulating Fock state of localized pairs forms in the center of the trap, displacing 
the superconducting state toward the boundaries.
When the spin population is imbalanced, two different limits are realized:
At low densities we have found a structure with two regions, a well defined superconducting inner core, and fully polarized metallic wings 
that effectively behave as a 1D gas of non-interacting spinless fermions. 
As the density increases, four phases emerge as a function of $S^z$: 
a Fock state in the center of the trap is gradually replaced by free $\downarrow$-fermions moving in a fully filled $\uparrow$ background. A superconducting shell 
separates it from the fully polarized wings. 
At finite magnetizations, the superconducting state becomes 
partially polarized, and can be described as a FFLO state with an oscillating order parameter 
and tightly bound pairs with a finite center-of-mass momentum.
 In the mean-field theory, the FFLO phase is regarded to be stable in a very narrow window around the Fermi surface,\cite{Leo AOP07}
 but we find that the presence of the trap helps to stabilize it by phase-separating the system.


We thank A. Moreo, D.J. Scalapino, and M. Troyer for helpful discussions, and S. Trebst for a critical reading of the manuscript. We are grateful to the 
Kavli Institute for Theoretical Physics at UCSB, where the idea for this work was conceived, and The Aspen Center for Physics for their hospitality.
F.H.-M.'s work   is supported in part by  NSF grant DMR-0706020, and by 
contract 
DE-AC05-00OR22725 with 
UT-Battelle, LLC.

\end{document}